\documentclass[11pt]{article}
\usepackage{amsmath,amssymb,color,epsfig,cite}
\usepackage{xcolor}
\usepackage{mathrsfs}
\usepackage{xcolor}

\usepackage{mathtools}

\usepackage{amsfonts}
\def\td{\tilde}
\newcommand{\hoch}[1]{$\, ^{#1}$}

\makeatletter
\@addtoreset{equation}{section}
\makeatother

\newcommand{\be}{\begin{equation}}
\newcommand{\ee}{\end{equation}}
\newcommand{\bea} {\begin{eqnarray}}
\newcommand{\eea}{\end{eqnarray}}
\newcommand{\nn}{\nonumber}

\def\ft#1#2{{\textstyle{\frac{\scriptstyle #1}{\scriptstyle #2} } }}
\def\fft#1#2{{\frac{#1}{#2}}}
\def\dfft#1#2{{\displaystyle\fft{#1}{#2}}}

\def\0{{\sst{(0)}}}
\def\1{{\sst{(1)}}}
\def\2{{\sst{(2)}}}
\def\3{{\sst{(3)}}}
\def\4{{\sst{(4)}}}
\def\5{{\sst{(5)}}}
\def\6{{\sst{(6)}}}
\def\7{{\sst{(7)}}}
\def\8{{\sst{(8)}}}
\def\sst#1{{\scriptscriptstyle #1}}
\def\oneone{\rlap 1\mkern4mu{\rm l}}
\def\ep{{\epsilon}}
\def\del{{\partial}}

\def\cF{{{\cal F}}}

\def\im{{{\rm i\,}}}
\def\R{{\mathbb R}}

\def\ie{{ i.e.~}}

\def\ben{\begin{equation}}
\def\bea{\begin{eqnarray}}
\def\een{\end{equation}}
\def\eea{\end{eqnarray}}

\def\ft#1#2{{\textstyle{\frac{\scriptstyle #1}{\scriptstyle #2} } }}
\def\fft#1#2{{\frac{#1}{#2}}}

\begin{document}

\begin{flushright}
\hfill {UPR-1333-T\ \ \ MI-HET-849}\\
\end{flushright}

\begin{center}

{\large {\bf 
Mass and Force Relations for Extremal EMDA Black Holes
}}

\vspace{15pt}
{\large S. Cremonini$^{1}$, M. Cveti\v c$^{2,3,4}$, 
              C.N. Pope$^{5,6}$ and A. Saha$^{7}$}

\vspace{15pt}

{\hoch{1}}{\it Department of Physics, Lehigh University, Bethlehem, 
PA 18018, USA}

{\hoch{2}}{\it Department of Physics and Astronomy,
University of Pennsylvania, \\
Philadelphia, PA 19104, USA}

{\hoch{3}}{\it Department of Mathematics, University of Pennsylvania, Philadelphia, PA 19104, USA}

{\hoch{4}}{\it Center for Applied Mathematics and Theoretical Physics,\\
University of Maribor, SI2000 Maribor, Slovenia}


\hoch{5}{\it George P. \& Cynthia Woods Mitchell  Institute
for Fundamental Physics and Astronomy,\\
Texas A\&M University, College Station, TX 77843, USA}

\hoch{6}{\it DAMTP, Centre for Mathematical Sciences,
 Cambridge University,\\  Wilberforce Road, Cambridge CB3 OWA, UK}

 \hoch{7}{\it Feza Gursey Center for Physics and Mathematics, Bogazici University, Kandilli 34684, Istanbul, T\"urkiye}

\vspace{10pt}

\end{center}

\begin{abstract}

We investigate various properties of extremal dyonic static black holes in 
Einstein-Maxwell-Dilaton-Axion theory.
We obtain a simple first-order
ordinary differential equation for 
the black hole mass in terms of its electric and
magnetic charges, which we can solve explicitly for certain 
special values of the scalar couplings.
For one such case we also construct new dyonic 
 black hole solutions, making use of the presence of an 
enhanced $SL(2,\mathbb{R})$ symmetry.
Finally, we investigate the structure of long range forces and binding energies between non-equivalent extremal black holes. 
For certain special cases, we can identify regions of parameter space where the force is always attractive or repulsive. 
Unlike in the case without an axion, the force and binding energies between distinct black holes are not always correlated with each other. 
Our work is motivated in part by the question of whether long range forces between non-identical states can potentially encode information about UV constraints on low-energy physics.

\end{abstract}

{\scriptsize 
cremonini@lehigh.edu, cvetic@physics.upenn.edu,
pope@physics.tamu.edu, aritra.saha@pt.bogazici.edu.tr.}

\pagebreak

\tableofcontents
\newpage
\section{Introduction}

A central question in the efforts to identify the signatures of quantum gravity on low-energy physics is to quantify what it means for gravity to be the weakest force.
To date, the most developed attempt at doing so is the Weak Gravity Conjecture (WGC) \cite{Arkani-Hamed:2006emk}, which claims, roughly, that  
theories of quantum gravity must contain 
states whose charge is greater than or equal to their mass.
For two such (identical) states, then, the gravitational attraction would clearly be weaker than (or at best balanced by) the electromagnetic repulsion. 
A natural extension of this idea is to examine the effects of all long range interactions acting on the states in the theory, including those mediated by massless scalar fields.
This has led to the 
 formulation of the Repulsive Force Conjecture (RFC) \cite{Arkani-Hamed:2006emk,Palti:2017elp,Heidenreich:2019zkl}, according to which theories of quantum gravity should contain states that are ``self-repulsive'', 
i.e. states which would feel either a repulsive or vanishing force when placed asymptotically far away from an identical copy of themselves. 
Both the WGC and the RFC lead to constraints on low energy effective field theories (EFTs) and have been explored in the context of the Swampland program  (see e.g. \cite{Harlow:2022ich} for a review of the WGC).
For implications of these conjectures for black holes in theories with higher derivatives, see e.g.  \cite{Cremonini:2021upd,Etheredge:2022rfl}.

Although the RFC is strictly a statement about interactions between identical objects, i.e.
self-forces,
it is natural to wonder whether forces between \emph{non-identical} black holes may also be somehow constrained by UV physics. If this is the case, studying their structure could help us shed light on the swampland. 
Indeed, this was part of the motivation behind our previous work \cite{Cremonini:2022sxf,crcvposa}, where we examined 
long range interactions between extremal dyonic black holes in Einstein-Maxwell-Dilaton (EMD) theories,
\bea
{\cal L}= R\, {*\oneone} - \ft12 {*d\phi}\wedge d\phi 
  - \ft12 e^{a\phi}\, {*F}\wedge F\, ,\label{emdlag}
\eea
and identified certain features which appeared to be generic. 
We found an interesting 
correlation between the range of the coupling 
$a$ parametrizing the strength of the dilaton coupling in the 
gauge field's kinetic term, the convexity/concavity of the surface 
characterizing the extremal mass as a function of the black hole charges, and 
the behavior of the long range forces among distinct solutions to the theory. 
In particular, we showed \cite{crcvposa} that in this class of EMD models, when $a>1$ 
the mass of each black hole solution is a \emph{convex} function of the black hole charges, 
 the long-range force (between non-identical black holes) is positive and so is the binding energy, hence the two black holes are likely to repel each other.  
On the other hand, when $a<1$  the mass function is \emph{concave} and the force attractive, associated with a negative binding energy. The special case $a=1$ then acts as a dividing line, with the forces between any two solutions always vanishing -- this is not surprising, since the corresponding extremal dyonic black holes are BPS. 
Given the generic results of \cite{crcvposa}, it is natural to ask whether one would find similar features in a theory that admits an axion, in addition to the dilatonic scalar.
While examining this question, we have also constructed new classes of solutions, as we discuss briefly below.

With these motivations in mind, in this paper we focus on 
extremal dyonic black hole solutions
to Einstein-Maxwell-Dilaton-Axion (EMDA) theory in 4 dimensions, 
whose Lagrangian is given by
\bea
\label{EMDALag}
{\cal L}= R\, {*\oneone} - \ft12 {*d\phi}\wedge d\phi - 
  \ft12 e^{-2 a \phi}\,{*d\chi}\wedge  d\chi - \ft12 e^{a\phi}\, {*F}\wedge F
  - \ft12 b\, \chi\, F\wedge F\, ,
\eea
where
the dilatonic and axionic couplings $a,b$ are taken to be arbitrary. 
The simpler EMD model (\ref{emdlag}) is recovered by taking 
$b=0$ and setting the axion to zero, $\chi=0$. 
For general values of the couplings, we do not expect to be able to
solve explicitly for the black hole solutions.
Indeed, this was also the case in the EMD theory, 
where only for certain values of $a$ (namely, $a=0, 1, \sqrt{3}$) 
explicit black hole metrics were known. 
As it turns out, for the EMDA case even less is known. 
The only known black hole solution now corresponds to $a=b=1$, which arises from a consistent truncation of the STU supergravity Lagrangian in 4 dimensions \cite{chowcomp}. 
Nonetheless, generalizing our analysis in \cite{crcvposa}, in this paper we 
are successfully able to 
construct a first order differential equation -- valid for arbitrary values 
of $a,b$ -- for a function $f(x)$ which encodes the mass and the other 
conserved charges of the black hole.  
Solving this differential equation, which we will refer to as the mass ODE, 
would in principle allow us to determine all the macroscopic properties of an extremal black hole for a given set of couplings -- the physical mass, electric and magnetic charges, as well as the dilatonic and axionic scalar charges supporting the solution.
However, we have been able to obtain explicit solutions for the mass function 
$f(x)$ only for special choices of the couplings, namely when $a=b$ and is 
arbitrary,  and when $a=1,\,  |b| \leq 1/\sqrt{2}$.  It should be
noted, however, that even in cases where the explicit solution for the
mass function $f(x)$ cannot be found, it is a rather straightforward matter
to solve for $f(x)$ numerically, thus encoding all the information about
the mass, electromagnetic charges and scalar charges in the single
function $f(x)$. 
In the special case $a=b$ we have also constructed  an explicit 
solution for the black hole metric,
thanks to the existence of an enhanced 
$SL(2,\mathbb{R})$ symmetry of the equations of motion in this case.

To determine whether some of the generic features seen in the EMD theory survive in the presence of an axion, we want to examine the behavior of long-range forces\footnote{The choice of the numerical factor in front of the mass term is purely due to convention and to match with our previous work.}
between two black holes in this theory,
\bea
\label{F12def}
F_{12} = \frac{1}{r^2} \left(Q_1Q_2 + P_1 P_2 - \ft14 M_1 M_2 - \Sigma_{\phi1} \Sigma_{\phi2} -   \Sigma_{\chi1} \Sigma_{\chi2}  \right),
\eea
where $ \Sigma_\phi$ and $\Sigma_\chi$ refer to the dilatonic and axionic scalar charges, respectively. What we are particularly interested in is whether we can identify ranges for the 
couplings that will ensure that the force is either attractive or repulsive, as was the case for the EMD theory. 
We also want to compare the long range interactions to the 
 binding energy 
between two extremal black holes, which we describe by
\bea
\label{DeltaMdef}
\Delta M = M(Q_1+Q_2,P_1+P_2) - M(Q_1,P_1) - M(Q_2,P_2)\, .
\eea
While in the EMD theory this was correlated precisely with the behavior of the force (as we would expect), we shall see that this is no longer the case here -- in the presence of an axion, the signs of the binding energy and force 
do not always agree with each other.
Moreover, among the instances when they do agree,  
there is one choice of couplings for which the force and binding energy vanish, despite the black holes being non-BPS.
We will return to this issue in the Conclusions.


The paper is organised as follows. 
Section \ref{Section2} lays out the fundamentals of the 
EMDA theory with general couplings and presents the derivation of the mass ODE. 
In Section \ref{Section3} we construct exact black hole solutions for the case $a=b$,
using the enhanced $SL(2,\mathbb{R})$ symmetry group, and discuss a perturbative solution for the mass function when the couplings are related via $a=b+\epsilon$, with $\epsilon <<1$.  
In Section \ref{Section4} we investigate two exact solutions to the mass ODE,  one for $a=b$ and the other for $a=1,\,  b^2 \leq 1/2$.
Section \ref{Section5} is devoted to the computation of the 
 long-range force and binding energy for specific choices of couplings.
We relegate closing comments to the Conclusions. In the Appendix we discuss a non-extremal generalisation of the solutions we constructed for the $a=b$ case.

\section{Einstein-Maxwell-Dilaton-Axion Theory}
\label{Section2}

\subsection{EMDA Lagrangian with arbitrary dilaton axion coupling}

In this section we describe the EMDA Lagrangian, discuss the asymptotic behavior of the black holes we are interested in and identify the constraint equation. 
This constraint equation, which comes from the combination of Einstein
equation components that involves only first derivatives of the metric 
functions, is then turned into a differential equation for the function 
$f(x)$ describing the  mass of the black hole solutions. 
Let us begin by considering the EMDA theory described by the Lagrangian
\bea
{\cal L}= R\, {*\oneone} - \ft12 {*d\phi}\wedge d\phi - 
  \ft12 e^{-2 a \phi}\,{*d\chi}\wedge  d\chi - \ft12 e^{a\phi}\, {*F}\wedge F
  - \ft12 b\, \chi\, F\wedge F\,,\label{emdalag}
\eea
where $a$ and $b$ denote the coupling constants, which are real.  
The relative factor in the exponents of the dilaton appearing in the
kinetic terms for the gauge field and for the axion field are
chosen so that the Lagrangian has in total 
an $\R\times \R$ global symmetry under which
\bea 
\phi &\longrightarrow& \phi' = \phi + k\,,\nn\\
\chi &\longrightarrow& \chi'= e^{a k}\, \chi + \lambda\,,\nn\\
F  &\longrightarrow &  F' = e^{-\ft12 a k}\, F\,.\label{scalings}
\eea
(The particular choice of dilaton coupling strengths implies the existence of 
the second $\R$ factor, associated with the dilaton shift symmetry $\phi
\longrightarrow \phi + k$, in addition to the generic $\R$ factor associated
with the axion shift symmetry $\chi\longrightarrow \chi+\lambda$.)  
The equations of motion for the theory (\ref{emdalag}) are given by 
\bea
d{*d}\phi - a e^{-2a\phi}\, {*d\chi}\wedge d\chi + 
         \ft12 a\, e^{a\phi}\, {*F}\wedge F &=&0\,,\nn\\
d(e^{-2a\phi}\, {*d\chi}) + \ft12 b\, F\wedge F &=&0\,,\nn\\
d(e^{a\phi}\, {*F}) + b\, d(\chi\, F) &=&0\,,\nn\\
R_{\mu\nu} -\ft12 \del_\mu\phi\, \del_\nu\phi -
   \ft12 e^{-2a\phi}\, \del_\mu\chi\, \del_\nu\chi  
   -\ft12e^{a\phi}\, (F^2_{\mu\nu} -\ft14 F^2\, g_{\mu\nu}) &=&0\,.
\eea
The conserved electric charge $Q$ (following from the equation of motion for
$F$) and magnetic charge $P$ are given by
\bea
Q &=& \fft1{4\pi} \int\big(e^{a\phi}\, {*F} + b\, \chi\, F\big)\,,\nn\\
P &=& \fft1{4\pi} \int F\,.
\eea
Under the transformations in eqns (\ref{scalings}), the charges will then transform as
\bea
Q&\longrightarrow& Q'= e^{\ft12 a k}\, Q + 
                b\, \lambda\, e^{-\ft12 a k}\, P\,,\nn\\
P&\longrightarrow&  P' = e^{-\ft12 a k}\, P\,.
\eea

\subsection{Static extremal dyonic black hole solutions}

  We shall study static spherically symmetric 
dyonic extremal black holes in this EMDA theory. The
metric may be taken to have the form
\bea
ds^2 = -e^{2U}\, dt^2 + e^{-2U}\, \big(dr^2 + r^2\, d\Omega^2\big)\,,
\eea
with the gauge potential taking the form
\bea
A= \psi(r)\, dt - P\, \cos\theta\, d\varphi\,.
\eea
The functions $U$ and $\psi$, together with the dilaton $\phi$ and
the axion $\chi$, will depend only of the radial variable $r$.
   If we assume that these functions all tend to zero at infinity, 
we may consider a large-$r$ expansion of the form
\bea
U &=& \fft{u_1}{r} + \fft{u_2}{r^2} + \fft{u_3}{r^3} +\cdots\,,\nn\\
\psi &=& \fft{\psi_1}{r} + \fft{\psi_2}{r^2} + 
         \fft{\psi_3}{r^3} +\cdots\,,\nn\\
\phi &=& \fft{\phi_1}{r} + \fft{\phi_2}{r^2} + 
        \fft{\phi_3}{r^3} +\cdots\,,\nn\\
\chi &=& \fft{\chi_1}{r} + \fft{\chi_2}{r^2} + 
\fft{\chi_3}{r^3} +\cdots\,.\label{largerexp}
\eea
Substituting the expansions into the equations of motion, one can solve for all the 
coefficients $(u_i,\psi_i,\phi_i,\chi_i)$ with $i\ge2$ in terms
of $(u_1,\psi_1,\phi_1,\chi_1)$ and the magnetic charge parameter $P$,
 together with the constraint
\bea
P^2 + \psi_1^2 - 4 u_1^2 -\phi_1^2-\chi_1^2=0\,.\label{constraint0}
\eea
The parameter $\psi_1$ can be identified 
with the electric charge $Q$, while 
$\phi_1$ and $\chi_1$ with, respectively, the dilatonic and axion scalar charges.
Thus we have
\bea
U=\fft{u_1}{r} + \cdots\,,\quad \psi=\fft{Q}{r}+\cdots\,,\quad 
\phi= \fft{\Sigma_\phi}{r}+\cdots\,,\quad
  \chi= \fft{\Sigma_\chi}{r}+\cdots\,.
\eea
With the metric taking the asymptotic form 
\bea
ds^2 &=& -\big(1+\fft{2u_1}{r}\big)\,dt^2 + \big(1-\fft{2u_1}{r}\big)\, 
(dr^2+ r^2\, d\Omega^2) +\cdots\,,\nn\\
&=& -\big(1+\fft{2u_1}{r}\big)\,dt^2 + h_{ij}\, dx^i\, dx^j+\cdots\,,
\eea
the mass may be calculated using the standard ADM formula
\bea
M_{\rm ADM} = \fft1{16\pi}\int d\Sigma_i\, (\del_j\, h_{ij} - \del_i\, h_{jj}) \, , 
\eea
in the limit when $r$ goes to infinity.  Thus, we have
\bea
M_{\rm ADM} = -u_1\,.
\eea
Using a convention where we define the mass to be $4 M_\text{ADM}$ to make contact with the previous literature \cite{crcvposa,heid}, the
constraint (\ref{constraint0}) becomes
\bea
Q^2+P^2 - \ft14 M^2 - {\Sigma_\phi^2} -{\Sigma_\chi^2}=0\,.\label{constraint}
\eea
This condition is in fact equivalent to the statement that there
is no force between two identical, widely-separated extremal black holes.

Note that 
(\ref{constraint}), which we have derived here
simply by solving the equations resulting from substituting the
large-$r$ expansions (\ref{largerexp}) into the equations of
motion, applies to any solutions at all of the assumed asymptotic
form.  As noted above, these constitute a four-parameter family of
solutions, characterised by the constants $(u_1,\psi_1,\phi_1,\chi_1,P)$ 
subject to the constraint (\ref{constraint0}).  Among these solutions, only a two-parameter
family are expected to describe genuine extremal black holes, which 
should be characterised by the two charges $P$ and $Q=\psi_1$.  The
general four-parameter asymptotic solutions will generically give
singular solutions in the interior when one numerically integrates
inwards.  One could use a shooting method to sift through the four-parameter
family of initial large-$r$ starting data, in order to find choices
that integrate in to a smooth extremal horizon behaviour.

\subsection{An equation for the mass}

   In the asymptotic form of the solution we considered above,
we assumed that the dilaton and axion went to zero at infinity.  We can
easily introduce non-zero asymptotic values $\phi_\infty$ and
$\chi_\infty$ by making use of the global symmetries described in
eqns (\ref{scalings}), by defining $k=\phi_\infty$ and
$\lambda=\chi_\infty$.  In terms of these redefined fields
we  have
\bea
\phi=\phi_\infty + \fft{\Sigma_\phi}{r} + \cdots\,,\qquad
\chi= \chi_\infty + \fft{\Sigma_\chi}{r}+\cdots\,,
\eea
with the charges now given by
\bea
Q &=& \fft{e^{\ft12 a \phi_\infty}}{4\pi} \int\big(e^{a\phi}\, {*F} + b\, \chi\, F\big)+ 
\fft{b\, \chi_{\infty}\, e^{-\ft12 a \phi_\infty}}{4\pi}\int F \,,\nn\\
P &=& \fft{e^{-\ft12 a\,\phi_\infty}}{4\pi} \int F\,.\label{chargesgen}
\eea
It was shown in \cite{gibkalkol} that by applying a Wald-type argument one
can derive contributions in the first law of thermodynamics of the form
$dM = \Sigma_\phi\, d\phi_\infty + \Sigma_\chi\, d\chi_\infty+\cdots$, and
hence that the scalar charges $\Sigma_\phi$ and $\Sigma_\chi$ are given by
\bea
\Sigma_\phi = \fft{\del M}{\del\phi_\infty}\,,\qquad
\Sigma_\chi = \fft{\del M}{\del\chi_\infty}\,.
\eea

   We know that the mass of an extremal dyonic black hole solution
must be expressible as a function solely of the electric and magnetic
charges, so
\bea
M= \cF(Q,P)\,.
\eea
Furthermore, on dimensional grounds it must be that $\cF(Q,P)$ is homogeneous of degree one, in the
sense that
\bea
\cF(\lambda Q,\lambda P)= \lambda\, \cF(Q,P)\label{homo}
\eea
for any constant $\lambda$.  By the chain rule we shall have 
\bea
\fft{\del}{\del \phi_\infty} = 
  \fft{\del Q}{\del\phi_\infty} \, \fft{\del}{\del Q} +
\fft{\del P}{\del\phi_\infty} \, \fft{\del}{\del P}\,,\qquad
\fft{\del}{\del \chi_\infty} =
  \fft{\del Q}{\del\chi_\infty} \, \fft{\del}{\del Q} +
\fft{\del P}{\del\chi_\infty} \, \fft{\del}{\del P}\,.
\eea
In view of the expressions (\ref{chargesgen}) for the electric and
magnetic charges we therefore have
\bea
\fft{\del}{\del \phi_\infty} =\ft12 a\,\Big(Q\, \fft{\del}{\del  Q} -
  P\, \fft{\del}{\del P}\Big)\,,
\qquad
\fft{\del}{\del \chi_\infty} = b\, P\, \fft{\del}{\del Q}\,.
\label{scaldir}
\eea
Note that, after having made use of the non-vanishing 
asymptotic values $\phi_\infty$ and $\chi_\infty$ in order to derive
these results, we then specialise back for simplicity to the case 
where $\phi_\infty$ and $\chi_\infty$ vanish.

   Using the expressions (\ref{scaldir}), it is now evident that the 
constraint equation (\ref{constraint}) implies that the mass function 
$\cF(Q,P)$ must satisfy the differential equation
\bea
Q^2+P^2 -\ft14 \cF^2  -\ft14 a^2\, 
  \Big( Q\, \fft{\del \cF}{\del Q} - P\, \fft{\del \cF}{\del P}\Big)^2 
-b^2\, P^2\, \Big(\fft{\del \cF}{\del Q}\Big)^2 =0\,.\label{masspde}
\eea
Differentiating the homogeneity relation (\ref{homo}) with respect to
$\lambda$, and then setting $\lambda=1$, implies that
\bea
Q\, \fft{\del\cF}{\del  Q} +
  P\, \fft{\del\cF}{\del P}\ =\cF\,.
\eea
This can be used in order to rewrite eqn (\ref{masspde}) purely as
an ordinary differential equation in terms of derivatives with respect
to $Q$, with $P$ viewed as a fixed parameter.  Alternatively, one could
turn the equation into an ordinary differential equation in terms
of derivatives with respect to $P$, with $Q$ viewed as a fixed parameter.
In fact, as can easily be seen, if we define\footnote{It should be noted that
in all our analysis, we are taking the electric and magnetic charges 
$Q$ and $P$ to be non-negative.}
\bea
\cF(Q,P) = \sqrt{8PQ}\, f(x)\,, \label{defcF}
\eea
where $x$ is defined by
\bea
Q= e^{2 a x}\, P\,, \label{relQP}
\eea
then the equation (\ref{masspde}) is equivalent to the statement that
$f(x)$ must satisfy the ordinary differential equation
\bea
\boxed{
f^2 + {f'}^2 + \fft{b^2}{a^2}\, e^{-4 a x}\, 
(f' + a\, f)^2 =\cosh 2 a x
}
\label{feqn}
\eea
where a prime denotes a derivative with respect to $x$.
If we solve this equation for $f(x)$, we can then simply write the
mass of the extremal black hole as 
\bea
M= \cF(Q,P)= \sqrt{8 PQ}\, f(x)\,,\qquad x= \fft{1}{2a}\, \log\fft{Q}{P}\,.
\label{massform}
\eea
The dilaton and axion scalar charges can then be written as 
\bea
\Sigma_\phi = \sqrt{2P Q} \, f'(x)\,,\qquad
\Sigma_\chi= \fft{b\, \sqrt{2PQ}}{a}\, e^{-2ax}\, \left(a f(x)+ f'(x)\right)\,. \label{scalch}
\eea
If we take the limit $b \longrightarrow 0$, then \eqref{feqn} reduces to the case without axion found in \cite{crcvposa}, and $\Sigma_\chi$ vanishes. 
In this limit, setting $\chi=0$ is consistent with the EMDA equations of motion, hence we go back to the 
Einstein-Maxwell-Dilaton (EMD) case.

Finally, before concluding this section, let us present the formula for the long-range force between two \emph{non-identical} black holes written in terms of $f(x)$. Using the expressions for $M,\Sigma_\phi,\Sigma_\chi$ in terms of $f(x)$ and trading the variable $Q$ for $x$ using \eqref{relQP}, we can express the force between the two black holes as
\bea
r^2\,F_{12} &=& Q_1\, Q_2 + P_1\,P_2 -\ft14 M_1\, M_2 -\Sigma_{\phi1}\, \Sigma_{\phi2}
-\Sigma_{\chi1}\,\Sigma_{\chi2}\,,\nn\\
&=& 2P_1P_2\,e^{a(x_1+x_2)}\,\Big\{\cosh[a(x_1+x_2)] - f(x_1)\, f(x_2) - f'(x_1)\, f'(x_2) \nn\\
&&
- \fft{b^2}{a^2}\, e^{-2a(x_1+x_2)}\, 
[f'(x_1)+ a\, f(x_1)]\,[f'(x_2) + a\, f(x_2)] \Big\}\,.
\eea
If we take $x_1=x_2$, the expression will reduce to the self-force, which is zero by \eqref{feqn}. 
When $x_1\neq x_2$, we associate $F_{12}>0$ with the force being ``repulsive,'' while $F_{12}<0$  with the force being ``attractive''.

\subsection{Small-$x$ expansion for $f(x)$}

    The mass equation (\ref{feqn}) generalises the equation 
$f^2 + {f'}^2 = \cosh 2a x$ that we derived previously in \cite{crcvposa} when studying static 
extremal dyonic black holes in the simpler EMD theory.  In that case the equation had a symmetry under sending $x\longrightarrow -x$,
implying that it would admit a solution for which $f(x)=f(-x)$, and indeed
the solution with this symmetry was the one that properly described the
mass of the black holes.  In the present case of black holes in the
EMDA theory, the situation is less clear.   Some insight can be
achieved by considering the solution for $f(x)$ in a  small-$x$ expansion.
   If we write 
\bea
f(x) = c_0 + c_1\, x + c_2\, x^2 + c_3\, x^3+\cdots\,,
\eea
then substituting into eqn (\ref{feqn}) implies, at leading order,
\bea
c_0^2 + c_1^2 + \fft{b^2\, (c_1 + a\, c_0)^2}{a^2} -1=0\,,\label{eqn0}
\eea
with the equations at higher order allowing one to solve linearly
for $c_2$, $c_3$, etc.  A natural choice, in the spirit of the 
symmetric solution that can be found in the EMD case, is to take
$c_1=0$ so that at least one has $f'(0)=0$.  Doing this, eqn
(\ref{eqn0}) implies that
\bea
c_0 = \fft{1}{\sqrt{1+b^2}}\,,
\eea
and then at higher orders we find 
\bea
c_2 = \fft{a^2}{\sqrt{1+b^2}}\,,\quad 
c_3=\fft{2 a^3\, (b^2-a^2)}{3b^2\, \sqrt{1+b^2}}\,,\quad
c_4= \fft{a^4\, (6a^4 -11 a^2 \, b^2 + 4 b^4 +a^2-b^2)}{6 b^4\,
   \sqrt{1+b^2}}\,, 
\eea
and so on.  It is evident that even if we choose the solution with $c_1=0$,
the higher odd-order powers of $x$ will appear in the series expansion.
Thus we cannot, in general, require $f(x)=f(-x)$.  Nevertheless, 
it does seem reasonable to impose the
requirement that the function $f(x)$ should obey $f'(0)=0$, in which case  the solution will also obey $f(0)= \fft{1}{\sqrt{1+b^2}}$.

\section{Exact Mass Solution For Arbitrary $a=b$}
\label{Section3}

\subsection{Generalisation of the Gibbons-Maeda black holes in EMDA theory}
We now embark on the task of finding solutions to the mass differential equation \eqref{feqn}. 
Although it is a first order ODE, it is non-linear, hence finding solutions for general values of $a,b$ is hard. 
However, it turns out that one can find solutions for specific choices of the couplings. One such choice corresponds to the special case $a=b$, which we discuss now. 
As we shall see, in this case not only can we find a solution to the mass ODE, but we can also obtain an exact solution for a static dyonic 
black hole 
for arbitrary values of $a$.  The reason for
this is that when $b=a$ the field equations  have a full $SL(2,\R)$ symmetry.
To see this, we note that when $b=a$, after introducing  rescaled fields
$\tilde\phi$, $\widetilde\chi$ and $\widetilde A$ such that
\bea
\tilde\phi=a\,\phi\,,\qquad \widetilde\chi= a\,\chi\,,\qquad
\widetilde A=a\, A\,,
\eea
the Lagrangian (\ref{emdalag})
becomes
\bea
{\cal L}= R\, {*\oneone} + \fft1{a^2}\, \Big\{
  - \ft12 {*d\tilde\phi}\wedge d\tilde\phi -
  \ft12 e^{-2 \tilde\phi}\,{*d\widetilde\chi}\wedge  d\widetilde\chi 
 - \ft12 e^{\tilde\phi}\, {*\widetilde F}\wedge \widetilde F
  - \ft12 \, \widetilde \chi\, \widetilde F\wedge 
    \widetilde F\Big\}\,.\label{temdalag}
\eea
With the exception of the overall $a^{-2}$ coefficient multiplying the
entire set of matter terms, this is otherwise just
like the $a=b=1$ Lagrangian, for which the field equations do 
have a full $SL(2,\R)$ symmetry. This is an enlargement of
the $\R\times\R$ global symmetry that arises for all the theories
we are considering, generically with $b\ne a$.
Since the metric is inert under the $SL(2,\R)$, this means that
the field equations following from the Lagrangian 
(\ref{temdalag}) have this symmetry too.  In addition, note that setting $b=a$ in
the original Lagrangian (\ref{emdalag}) was essential in order
to get the correct coefficient of the $\widetilde\chi\,\widetilde F\wedge
\widetilde F$ term in (\ref{temdalag}), such that the theory has the
full $SL(2,\R)$ symmetry.

Using this symmetry, and in particular its $U(1)$ subgroup, we can start from
the purely electric or purely magnetic static black hole
solution that was found by Gibbons and Maeda in the EMD theory
for arbitrary $a$. 
Such black holes are also solutions of the
EMDA theory, since for purely electric or purely magnetic charges there are no contributions from the $\chi\, F\wedge F$
term, and therefore $\chi$ can be consistently truncated to zero.  We 
then act on the Gibbons-Maeda solution with the
$U(1)$ global symmetry transformation. This rotates it into
a solution with both electric and magnetic charges, while at the same time turning on the axion field $\chi$.
This procedure constructs for us
the desired dyonic black hole solutions of the $b=a$ EMDA theory for
arbitrary values of $a$.\footnote{It should be noted that although the
EMDA theories with $b=a$ all have a full $SL(2,\R)$ global symmetry, 
if we reduce the theory on the time direction to three dimensions
we do {\it not}, in general, get a theory with an enhanced global
symmetry group that could be used to generate charged solutions
from uncharged ones.  In other words, we would merely get a theory
with $SL(2,\R)\times SL(2,\R)\times \R\times \R$ global symmetry.
This would comprise the 
$SL(2,\R)$ factor already present in $D=4$, an additional
$SL(2,\R)$ factor coming from the reduction of $D=4$ Einstein gravity
to $D=3$ Einstein gravity, and two $\R$ factors coming from the
shift symmetries of the two axions arising from the reduction of
the $D=4$ Maxwell field.  It is only for $a=1$ that a symmetry enhancement
to $O(3,2)$ would occur.  (This would happen because the 
dilaton vectors
of the two extra axions from the reduction of the Maxwell field would
have the correct lengths to become the additional two positive
root vectors of $O(3,2)$.  (See \cite{cjlp,cjlp3} for a discussion
of global symmetry enhancements in toroidal dimensional reductions.))}

   It is evident from the Lagrangian (\ref{temdalag}) that the scalar
sector will have an $SL(2,\R)$ global symmetry that is implemented by
defining the complex field\footnote{Note that our dilaton $\tilde\phi$ has
the opposite sign to the conventional one for an axion/dilaton system,
hence the $e^{+\tilde\phi}$ in eqn (\ref{taudef}).}
\bea
\tau = \widetilde\chi + \im\, e^{\tilde\phi}\,,\label{taudef}
\eea
and then making fractional linear transformations under which
\bea
\tau\longrightarrow \fft{{\td a}\,\tau+{\td b}}{{\td c}\,\tau+{\td d}}\, ,
\eea
with ${\td a}$, ${\td b}$, ${\td c}$ and ${\td d}$ real constants 
which obey ${\td a}\, {\td d}-{\td b}\, {\td c}=1$.
(The ${\td a}$ and ${\td b}$ here are not to be confused with the 
$a$ and $b$ parameters in the EMDA Lagrangian (\ref{emdalag})!)  

  We take as our starting point the static black hole solutions in EMD
theory with arbitrary dilaton coupling $a$, as obtained by Gibbons and
Maeda \cite{gibmae}.  
In a convenient presentation given in \cite{cvegibpop}, using
an isotropic radial coordinate $\rho$, these solutions are given by
\bea
ds^2 &=& -\Delta\, dt^2 + \Phi^4\, (d\rho^2 + \rho^2\, d\Omega^2)\,,\nn\\
A &=& 2P\,\cos\theta\, d\varphi\,,\qquad e^{a\phi} = 
f_-^{\fft{2a^2}{1+a^2}} \,,\nn\\
f_\pm &=& \fft{\Big(1\mp \fft{\beta\gamma}{\rho}\Big)^2}{\Big(1+\fft{\beta^2}{\rho}\Big)
\Big(1+\fft{\gamma^2}{\rho}\Big)}\,,\nn\\
\Delta&=& f_+\, f_-^{\fft{1-a^2}{1+a^2}}\, F_\pm\,,\qquad
\Phi^2= \Big[\Big(1+\fft{\beta^2}{\rho}\Big)
   \Big(1+\fft{\gamma^2}{\rho}\Big)\Big]^{\fft1{1+a^2}}\,
  \Big(1+\fft{\beta\gamma}{\rho}\Big)^{\fft{2a^2}{1+a^2}}\,, \label{GibMae1}
\eea
where $\beta$ and $\gamma$ are constants related to the mass and charge.  We have
adapted some of the notation and conventions in \cite{cvegibpop} to
ours.  In this 
presentation of the Gibbons-Maeda solution the black hole carries just a
magnetic charge $P$, related to the constants $\beta$ and $\gamma$ by
\bea
P^2 = \fft{\beta^2-\gamma^2}{1+a^2}\,.
\eea
The extremal limit is obtained by setting $\gamma=0$.  The solution then
becomes
\bea
ds^2 &=& e^{2U}\, dt^2 + e^{-2U}\, (d\rho^2 + \rho^2\, d\Omega^2)\,,\nn\\
e^{a\phi} &=& h\,,\qquad A=2P\cos\theta\, d\varphi\,,
\eea
where we have defined
\bea 
h= \Big(1+ \fft{\alpha}{\rho}\Big)^{-\fft{2a^2}{1+a^2}}\,,\qquad \hbox{and}
\quad \alpha= \ft12 \sqrt{1+a^2}\, P\,.
\eea

  We are now ready to apply a $U(1)\subset SL(2,\R)$ transformation in order
to obtain a dyonic black hole solution in the EMDA theory with $b=a$.
The $U(1)$ subgroup corresponds to fractional linear transformations of
the complex field $\tau$ with the $SL(2,\R)$ matrix
\bea
\begin{pmatrix} {\td a} & {\td b}\\ {\td c} & {\td d}\end{pmatrix}=
 \begin{pmatrix} c & s\\ -s & c \end{pmatrix}\,,\qquad
\hbox{where}\quad c=\cos\beta\,,\quad s=\sin\beta\,.
\eea
Since our starting point is the Gibbons-Maeda solution for which the
axion is zero, this leads, after the rotation, to
\bea
e^{\tilde\phi}= 
\fft{e^{\tilde\phi_0}}{c^2+ s^2\, e^{2\tilde\phi_0}}\,,\qquad
\widetilde\chi = \fft{s \,c\, (1-e^{2\tilde\phi_0})}{
c^2+ s^2\, e^{2\tilde\phi_0}}\,,
\eea
where $\tilde\phi_0$ denotes the dilaton field before the rotation, and
$c=\cos\beta$, $s=\sin\beta$.
The original purely magnetic charge $P_0$ will gives rise to the
rotated charges
\bea
P= c\, P_0\,,\qquad Q= -s\, P_0\,,
\eea
with the angle $\beta$ related to $Q$ and $P$ by
\bea
c=\cos\beta= \fft{P}{\sqrt{P^2+Q^2}}\,,\qquad s=\sin\beta=-
  \fft{Q}{\sqrt{P^2+Q^2}}\,.
\eea

After converting back to the original unrescaled dilaton and axion fields
$\phi$ and $\chi$, the dyonic solution therefore becomes
\bea
ds^2 &=& -e^{2U}\, dt^2 + e^{-2U}\, (d\rho^2+\rho^2\, d\Omega^2)\,,\qquad
e^{-2U}=\Big(1+\fft{\alpha}{\rho}\Big)^{\fft2{1+a^2}}\,,\nn\\
e^{a\phi} &=& \fft{h\, (P^2+Q^2)}{P^2 + h^2\, Q^2}\,,\qquad
\chi= -\fft{P Q\, (1-h^2)}{P^2+h^2\, Q^2}\,,
\eea
where
\bea
h= \Big(1+\fft{\alpha}{\rho}\Big)^{-\fft{2a^2}{1+a^2}}\,,\qquad
\alpha= \ft12\sqrt{1+a^2}\, \sqrt{P^2+Q^2}\,.
\eea
The gauge potential is given by
\bea
A = \fft{Q}{\rho+\alpha}\, dt + P \cos\theta\, d\varphi\,.
\eea
The mass, obtained by rescaling the ADM mass by a factor of 4 as in our
previous discussion, is given by
\bea
M= \fft{4\alpha}{1+a^2} = \fft{2}{\sqrt{1+a^2}}\, \sqrt{P^2+Q^2}\,. \label{massaeqb}
\eea
Finally, from the mass we can read off the
function $f(x)$ given in eqn (\ref{massform}), finding
\bea
f(x) = \fft1{\sqrt{1+a^2}}\, \big(\cosh 2 a x\big)^{\ft12}\,.\label{fxres}
\eea
It can be verified this this function indeed satisfies eqn (\ref{feqn})
when $b$ is set equal to $a$. Note that the solution also satisfies the boundary condition $f'(0)=0$, since it is symmetric in $x$.

It turns out to be convenient, as we learned in \cite{crcvposa}, to define the quantities
\bea
 u= \fft{Q}{M}\,,\qquad v = \fft{P}{M}\,,
\eea
which can also be expressed as
\bea
u(x)= \fft{e^{ax}}{\sqrt8\, f(x)}\,,\qquad v(x)= \fft{e^{-ax}}{\sqrt8\, f(x)}
\,.\label{uvres}
\eea
Since the mass is given by $M=\cF(Q,P)$, and $\cF(Q,P)$ obeys the
homogeneity condition $\cF(\lambda Q,\lambda P)=\lambda\, \cF(Q,P)$ as
in eqn (\ref{homo}), it follows that the extremal mass condition 
corresponds to the curve $\cF(u,v)=1$ in the $(u,v)$ plane. 

For the family of $b=a$ dyonic extremal black holes found above, which obey
eqn (\ref{fxres}), we therefore see that the extremal curve
$\cF(u,v)=1$, given parametrically by eqns (\ref{uvres}), is
the positive quadrant of the circle
\bea 
u^2+v^2= \ft14 (1+a^2)\,. \label{uvplota=b}
\eea
We should note that the black hole solution for the special case of $a=b=1$ can be obtained as a truncation of STU black holes, which are solutions of ${\cal N}=2$ supergravity coupled to three abelian vector multiplets in 4 dimensions \cite{chowcomp}. 
In this theory, there are three dilatons ($\phi_1,\phi_2,\phi_3$), three axions ($\chi_1,\chi_2,\chi_3$) and 4 abelian gauge fields ($A^1,{\td A}_2,{\td A}_3,A^4$). 
After performing the truncation $\phi_2=\phi_3=0$, $\chi_2=\chi_3=0$ and $A^1=A^4, {\td A}_2={\td A}_3$, followed by setting ${\td A}_3=0$, one obtains the EMDA Lagrangian with $a=b=1$. The black hole metric takes the following form
\bea
e^{-2U} &=& 1+\fft{\alpha}{r}\,,\nn\\
e^\phi &=& \fft{2(P^2+Q^2)\,r\, (r+\alpha)}{2 (P^2+Q^2)\, r^2 + 
  4 \alpha\, P^2\, r + P^2\, (P^2+Q^2)}\,,\nn\\
\chi &=& - \fft{\alpha\, PQ\,(P^2+Q^2 +4\alpha\, r)}{
  (P^2+Q^2)(2\alpha\, r^2 + 2P^2\, r + \alpha\, P^2)}\,, \nn\\
A &=& \fft{2 \alpha\, Q}{P^2+Q^2 + 2\alpha\, r}\,dt + P\,\cos\theta\,d\varphi\,,
\eea
where 
\bea
\alpha = \fft{\sqrt{P^2+Q^2}}{\sqrt2}\,.
\eea
The black hole mass simply becomes 
\bea
M=\sqrt{2}\,\sqrt{Q^2+P^2}\,,
\eea
which corresponds to 
\bea
f(x) = \ft1{\sqrt{2}}\,(\cosh 2x)^\ft12\,.
\eea
Finally let us note that one can also construct the EMDA generalisation of the \emph{non-extremal} Gibbons-Maeda black holes by reinstating the constant $\gamma$ (setting $\gamma=0$ corresponds to taking the extremal limit). Since it is not directly relevant to the discussion that follows, we present it in the Appendix.

\subsection{Perturbative solution for $f(x)$ when $ b = a + \epsilon$}

If we choose the couplings $a$ and $b$ 
to be perturbatively close to each other,  $b=a+\epsilon$ with $\epsilon<<1$, and define the perturbation for $f(x)$ around $b=a$ in the following way,
\bea
f(x) = \fft{1}{\sqrt{1+a^2}}\,(\cosh 2ax)^{1/2}\,\Big( 1 + \epsilon\,e^{-ax}\, w(x) + {\cal O}(\epsilon^2) \Big)\,,
\eea
we find, after plugging this ansatz into the mass ODE \eqref{feqn}, 
that at ${\cal O}(\epsilon)$ the function 
$w(x)$ satisfies the ordinary differential equation
\bea
w'(x) + \fft1a\,w(x) + \fft{e^{ax}}{(\cosh 2ax)^2} = 0\,.
\eea
This equation can be integrated by making use of the integrating factor $e^{x/a}$, giving
\bea
w(x) = e^{-\fft{x}{a}}\, \int e^{\fft{1+a^2}{a}\,{\td x}}\,(\cosh 2a{\td x})^{-2}\,d{\td x}\,,
\eea
which integrates to a hypergeometric function,
\bea
w(x) = e^{-\ft{x}{a}}\,c_1 + \fft{4a}{1+5a^2}\,e^{5ax}\, 
_2F_1\,\Big(2,\fft{5a^2+1}{4a^2}, \fft{9a^2+1}{4a^2}; - e^{4ax} \Big)\,,
\eea
where $c_1$ is a constant of integration. 

When $a=1$, the perturbation $w(x)$ takes a somewhat simpler form and becomes,
\bea
w(x) = c_1\,e^{-x} + e^{-x}\, \arctan e^{2x} - \fft{e^x}{1+e^{4x}}\,.
\eea
The constant $c_1$ can be easily determined by demanding that the perturbed 
function $f(x)$ will also obey the boundary condition $f'(0)=0$. 

Having an analytic expression for $f(x)$ when $b$ is slightly away from $a$, one can in principle determine how the macroscopic quantities (such as mass, electric and magnetic charge, scalar charges) for the dyonic Gibbons-Maeda black holes behave under the perturbation. Moreover using the perturbed $f(x)$, one can also determine how the binding energy and the force between two non-identical dyonic Gibbons-Maeda black holes will change under such perturbation. Although we do not carry out these computations explicitly for the dyonic Gibbons-Maeda case, we perform those for the case when $a$ is varied slightly away from one in section 4 and section 5.

\section{Exact Mass Solution For $a=1$}
\label{Section4}

\subsection{Expression for the mass when $a=1$ and $|b| \leq 1/\sqrt{2}$}

Another example where the mass function $f(x)$ admits an exact solution is when $a=1$ and $b$ lies in the range
\bea
-\fft1{\sqrt2} \le b \le \fft1{\sqrt2}\,.\label{brange}
\eea
To understand where this range for the axionic coupling comes from, consider
a trial solution of the form
\bea
f(x)= \cosh x - c\, e^{-x}\,.\label{fceqn}
\eea
Substituting this into the equation (\ref{feqn}) for the mass function and letting
$a=1$, we find that the equation is satisfied if
\bea
b^2 + 2c\, (c-1)=0\,,
\eea
which implies
\bea
c= \ft12 \pm\ft12 \sqrt{1-2b^2}\,.\label{cfromb}
\eea
Thus, we obtain a real solution for $f(x)$ provided that $b$ is in
the range given in eqn (\ref{brange}).
For general values of $b$ lying in the range (\ref{brange}), since we
have 
\bea
f(x) = \ft12 e^x + (\ft12 -c)\, e^{-x}\label{fa1}\, ,
\eea
we must require $c\le \ft12$ in order to have a mass function that is 
non-negative for all $x$.  In turn, this forces us to choose the minus sign
in eqn (\ref{cfromb}):
\bea
c= \ft12 -\ft12\sqrt{1-2b^2}\,.\label{cfromb2}
\eea
One interesting observation is that this solution violates the boundary 
condition which was
satisfied by the $a=b$ solution. Now we have
\bea
f'(0) = c \neq 0 \quad \text{for} \quad b \neq 0\,.
\eea
Thus within this class of solutions, 
only for $b=0$ (i.e. the EMD case with the choice $a=1$) does 
our $f(x)$ satisfy the boundary condition $f'(0)=0$.

It is instructive to convert the expression for $f(x)$ into one for the physical mass $M$. Using the relation \eqref{massform} we can write the mass as
\bea
M = {\cal F}(Q,P) = \sqrt{2}\,\Big(Q + \sqrt{1-2b^2}\,P \Big)\,. \label{massfora=1}
\eea
This is a particularly interesting result because $M$ is linear in $Q$ and 
$P$, which suggests a similarity with the dyonic BPS black hole solution of the $a=1$ EMD theory. In that case the mass was simply $M=\sqrt{2}\,(Q+P)$ (which is the $b=0$ limit of the above mass). 
However, the similarity is surprising because the black holes we are dealing 
with (for arbitrary $b$ in the range $-1/\sqrt{2} \leq b \leq 1/\sqrt{2}$) 
are perhaps not supersymmetric when $b\neq 0$.
Finally, note that the expression for the extremal curve ${\cal F}(u,v) = 1$, where $u=Q/M$ and $v=P/M$, now reduces to
\bea
\sqrt{2}\,(u + \sqrt{1-2b^2}\,\,v) = 1\,,
\eea
which is a straight line with vanishing extrinsic curvature, as expected from the linear form of the mass. This was also the case in the EMD theory, where the straight line was associated with the $a=1$ BPS solutions.

\subsection{Perturbing around the exact $a=1$ solution for $f(x)$}

Next, we will examine a perturbation of the $a=1$ solution (\ref{fceqn}) for $f(x)$.
Setting $a=1+\ep$ and working to first order  in $\ep$, we take the perturbed mass function $f(x)$
%
to have the form
\bea
f(x) = \big(\cosh a x - c e^{-a x}\big)  + \ep\, w(x)\,. \label{farounda=1}
\label{fpert}
\eea
Using eqn (\ref{cfromb2}) to express $b^2$ in terms of $c$, and substituting the expression above 
into the differential equation (\ref{feqn}), we obtain 
\bea
&&[e^{2x}\, (e^{4x}-1) + 4 c\,(2 e^{2x}+1) -4 c^2\, (2 e^{2x}+3) + 8 c^3]\,
w'(x) + 2 e^{2x}\, (e^{4x}+1)\, w(x) \nn\\
&&+ e^{2x}\, \big(e^{2x} + 2c-1\big)^2=0\,.
\label{weqn}
\eea
For general values of $c$ within the allowed range $0\le c\le\ft12$, 
the solution for $w(x)$ is quite complicated.  However, at the endpoints of 
the range, the form of $w$ is much simpler. The case $c=0$, 
corresponding to $b=0$, was already studied in our previous work \cite{crcvposa}.
At the other end of the range, namely $c=\ft12$, corresponding to $b^2=\ft12$, 
the solution to (\ref{weqn}) is given by
\bea
w(x)= \ft14 e^{-x}\, \arctan e^{2x} - \ft14 e^x + k\, e^{-x}\,, \label{wpert}
\eea
where $k$ is an arbitrary constant. Including this perturbation, $f(x)$ takes the following form for $a=1+\epsilon,\, b^2 = \ft12$ up to ${\cal O}(\epsilon)$
\bea
f(x) = \ft12\,e^x + \ft14\,e^{-x}\,(4\,k - e^{2x} + 2\,e^{2x}\,x + \arctan e^{2x}) \,\epsilon\,. \label{fpertb^2=1/2}
\eea
Note that when $x$ becomes large and negative ($P >> Q$), $f(x)$ will behave as $f(x) \sim k\,e^{-x}\,\epsilon$. Hence for non-zero $k$, depending on its sign, $f(x)$ can become negative for $\epsilon>0$ or $\epsilon<0$. Therefore, to ensure that the perturbed mass function stays positive for all $x$, we should choose $k=0$.

\section{Non-Identical Black Holes; Force And Binding Energy}
\label{Section5}

One of the main motivations for developing the formalism for expressing the extremal black hole mass in terms of $f(x)$ is to be able to calculate the long-range force between two non-identical black holes efficiently. We did so in the case of EMD black holes in our previous work \cite{crcvposa}, where we found a correlation between the sign of the force between two arbitrary black holes and their binding energy. 
In particular, we saw that when the dilaton coupling was $a>1$, both the force 
and the binding energy were positive ($F_{12}>0,\, \Delta M>0$).
On the other hand, for $a<1$ the force and the binding energy were both 
negative ($F_{12}<0,\, \Delta M<0$). The $a=1$ BPS solutions served as the 
dividing line, with the solutions experiencing a no-force condition and 
vanishing binding energy
($F_{12}=0,\, \Delta M=0$). While the latter is expected for BPS solutions, 
the correlation between the coupling and the attractive vs. repulsive 
nature of the long range forces was novel.

We now wish to investigate the same question in the presence of the axion. We shall do so in the two special cases we discussed in Sections \ref{Section3} and \ref{Section4}. First, when $a=b$ we will
compute the force and binding energy making use 
of the explicit form of $f(x)$ constructed in Section \ref{Section3}.  
We will then repeat the computations for the case  $a=1+\epsilon$ we discussed in Section \ref{Section4}, and investigate if the correlation between the force and the binding energy persists in the presence of the axion as we vary $a$ around $a=1$. For convenience let us quote the formulae for the force again,
\bea
r^2\,F_{12} &=& Q_1\, Q_2 + P_1\,P_2 -\ft14 M_1\, M_2 -\Sigma_{\phi1}\, \Sigma_{\phi2}
-\Sigma_{\chi1}\,\Sigma_{\chi2}\,,\nn\\
&=& 2P_1P_2\,e^{a(x_1+x_2)}\,\Big\{\cosh[a(x_1+x_2)] - f(x_1)\, f(x_2) - f'(x_1)\, f'(x_2) \nn\\
&&
- \fft{b^2}{a^2}\, e^{-2a(x_1+x_2)}\, 
[f'(x_1)+ a\, f(x_1)]\,[f'(x_2) + a\, f(x_2)] \Big\}\,. \label{F12} 
\eea
The expression for the binding energy is given as \cite{crcvposa}
\bea
\Delta M &=& {\cal F}(Q_1+Q_2,P_1+P_2) - {\cal F}(Q_1,P_1) - {\cal F}(Q_2,P_2) \,, \label{DelM}
\eea
where ${\cal F}(Q,P)$ is defined as ${\cal F}(Q,P) = \sqrt{8PQ}\,f(x) = \sqrt{8P}\,e^{a\,x}\,f(x)$ using \eqref{relQP}.

\subsection{Force between non-identical black holes when $a=b$}
We now calculate the force between two non-identical black holes for $a=b$. Plugging eqn (\ref{fxres}) into the expression for $F_{12}$ 
in eqn (\ref{F12}), we find
\bea
\fft{F_{12}}{P_1 P_2} = 2 e^{a(x_1+x_2)} \Big[\cosh[a (x_1 + x_2)] - \fft{\cosh2a x_1\, \cosh 2 a x_2 + a^2\,
  (1 + \sinh 2 a x_1\,\sinh 2 a x_2)}{(1+a^2)\, \sqrt{
\cosh 2 a x_1\, \cosh 2 a x_2}} \Big]\,. \nn\\ \label{forceaeqb}
\eea
As expected, this vanishes when $x_1=x_2$. By introducing the following 
changes of variable,
\bea
X_{1} = a\,x_{1}\,,\qquad X_2 = a\,x_2 \,, \qquad X_{\pm} = a\,(x_1 \pm  x_2)\,,
\eea
the force can be rewritten as
\bea
\fft{F_{12}}{P_1 P_2}= \fft{e^{X_+}\,(\cosh 2X_- - 1)}{\left(\cosh X_+ +\sqrt{\cosh 2X_1 \cosh 2X_2 }\right)} \left( \fft{a^2-1}{a^2+1} + \fft{2a^2}{a^2+1} W(X_1,X_2)\right)\,, \label{forceaeqb2}
\eea
where
\bea
W(X_1,X_2) = \fft{\cosh X_+}{\sqrt{\cosh 2X_1 \cosh 2X_2 }}.
\eea
To derive the result above we needed to employ the identity 
\bea
\fft{\cosh^2(X_+)}{\cosh(2X_1)\cosh(2X_2)}= 1+ \fft{1-\cosh(2X_-)}{2\cosh(2X_1)\cosh(2X_2)}\, . \label{ratio}
\eea
Using the expression  in \eqref{forceaeqb2}, we can determine whether the force is of a definite sign for a given $a$. To see this, note that the factor in front of the parenthesis is non-negative, and 
\bea
0 \leq W(X_1,X_2) \leq 1  \quad \text{for any} \quad X_1,X_2.
\eea
Therefore we can say the following,
\bea
 a^2 \geq 1\,, &&  F_{12} \geq 0\,, \quad \text{for any} \quad X_1,X_2\,, \nn\\
 \ft{1}{3} \leq a^2 < 1\,,  && F_{12} \geq 0 \quad \text{when} \quad W(X_1,X_2) \rightarrow 1\,; \quad F_{12} \leq 0 \quad \text{when} \quad W(X_1,X_2) \rightarrow 0 \,, \nn\\
0 < a^2 < \ft13\,, && F_{12} \leq 0\,, \quad  \text{for any} \quad X_1,X_2\,. 
\eea
In other words the force is positive (repulsive) when $a^2\geq 1$ and negative (attractive) when $0 < a^2 < \ft13$. When $\ft13 \leq a^2 < 1$ the force can change sign depending on $x_1,x_2$. In particular note that for $a^2=1$, which corresponds to the case which can be embedded inside the STU theory, the force is positive, or repulsive.

\subsection{Force between non-identical black holes when $a=1$ and $b^2<\ft12$}

Using the formula for $F_{12}$ in \eqref{F12} and the expression for $f(x)$, which for $a=1,\,b^2<1/2$ takes the form
\bea
f(x) = \cosh x - c\,e^{-x}\,, \quad c = \ft12 - \ft12\,\sqrt{1-2b^2}\,,
\eea
one can compute the force and show that
\bea
F_{12} = 0\,, \quad \text{when $a=1,\, \, b^2\leq \fft12$}.
\eea
Note that the vanishing of the force reinforces the similarity between the 
black holes for $a=1, b \neq 0$ (and $b^2\leq 1/2$), and the BPS black holes 
of the EMD theory for $a=1,b=0$. As is well known, vanishing of the force 
between two non-identical black holes is a feature which is commonly seen 
in BPS black holes. In the present case, the origin of the zero-force 
condition is less clear.

\subsection{Force between non-identical black holes when $a$ close to $1$, $ b^2=\ft12$}

Given that the $a=1$ solutions we examined above obey a zero-force condition 
independently of the value of the axionic coupling (as long as  
$b^2\leq \ft12$), it is of interest to ask what happens as we allow $a$ to 
vary slightly away from one. \eqref{weqn} captures the differential equation satisfied by the perturbation $w(x)$ to the mass function $f(x)$ when $a$ is varied slightly away from one, but $b$ is kept arbitrary in the range $|b| \leq 1/\sqrt{2}$. However, writing down a general solution for $w(x)$ is quite complicated, and we choose to focus on the case $b = 1/\sqrt{2}$ where \eqref{weqn} can be integrated easily. The perturbed $f(x)$ now takes a simpler form given in \eqref{fpertb^2=1/2} and can be used in \eqref{F12} to produce the force between two non-identical extremal black holes (while setting $a=1+\epsilon$ and expanding up to ${\cal O}(\epsilon)$)    
\bea
\fft{F_{12}}{P_1P_2} &=&  -2\,e^{(x_1+x_2)}\,\fft{ \sinh^2(x_1-x_2)\, \sinh(x_1+x_2)}{2\cosh2x_1\, \cosh2 x_2}\,\epsilon
+ {\cal O}(\ep^2)\,.
\eea
Now we see that up to order $\ep = a-1$, the sign of the force depends on the sign of $(x_1+x_2)$, due to the $\sinh(x_1+x_2)$ term in the numerator. 
In particular, for $a=1+\ep$ and $b^2=\ft12$,      if $x_1+x_2>0$ (\ie if $Q_1 Q_2>
P_1P_2$) the force between non-identical
extremal black holes will be 
positive
(that is, repulsive) if $a>1$ and negative (attractive) if $a<1$.
If, on the other hand, $x_1+x_2<0$ (\ie if $Q_1Q_2<P_1 P_2$), then the force 
will attractive when $a>1$ and repulsive when $a<1$. 
This can be summarized as follows:
\bea
&& \text{when} \quad x_1+x_2>0\,,  Q_1Q_2>P_1P_2\,, \quad   a>1\,, F_{12}<0\,, \quad a<1\,, F_{12}>0\,, \nn\\
&& \text{when} \quad x_1+x_2<0\,,  Q_1Q_2<P_1P_2\,, \quad    a>1\,, F_{12}>0\,, \quad   a<1\,, F_{12}<0\,. 
\eea
Thus, we see that for these solutions there is a sense in which $a=1$ (which corresponds to exactly no force) is a dividing line between attractive and repulsive interactions, with the caveat that the nature/sign of the interaction also depends on whether the electric charges are larger or smaller than the magnetic ones.  Even though we see some remnants of the behavior we observed in the simpler EMD case, here the overall structure is significantly more complicated.

\subsection{Binding energy between non-identical black holes when $a=b$}

Computing the binding energy for two arbitrary black holes for $a=b$ is straightforward if we use the expression for the physical mass given in \eqref{massaeqb}. Using the expression for binding energy in \eqref{DelM} and noting that the mass is given by,
\bea
M = {\cal F}(Q,P) = \fft{2}{\sqrt{1+a^2}}\,\sqrt{Q^2+P^2}\,,
\eea
the binding energy becomes,
\bea
\Delta M = \fft{2}{\sqrt{1+a^2}}\, \Big[ \sqrt{(P_1+P_2)^2 + (Q_1+Q_2)^2 } - \sqrt{P_1^2+Q_1^2} - \sqrt{P_2^2 + Q_2^2} \Big]\, .
\eea
Writing the latter as $\Delta M = \ft{2}{\sqrt{1+a^2}}(f_1 - f_2)$, one can easily show that,
\bea
f_1^2 - f_2^2 = 2P_1 P_2 + 2 Q_1 Q_2 - 2\sqrt{P_1^2P_2^2 + Q_1^2Q_2^2 + P_1^1Q_2^2 + Q_1^2P_2^2} \leq 0\,.
\eea
Thus, the binding energy is $\Delta M \leq 0$ for all $a=b$. 

This agrees with the shape of the extremal curve ${\cal F}(u,v)=1$, which in this case is given by $u^2+v^2=1$, i.e. since the curve is a convex circle in 
the $u,v$ plane, 
its extrinsic curvature is negative, which should agree with the sign of 
the binding energy by construction \cite{crcvposa}. However, unlike the EMD example, where the sign of the binding energy and the sign of the force correlated with each other, we see that this is not the case here. In fact only when $0<a^2<\ft13$, the sign of $F_{12}$ agrees with that of $\Delta M$. 
When $a^2\geq 1$, the signs are opposite.
Finally, in the intermediate range $\ft13 \leq a^2 <1$, $F_{12}$ may or may not agree with $\Delta M$ depending on $x_1,x_2$, i.e. the charges of the two initial black holes.

\subsection{Binding energy for $a=1,\, b^2\leq\ft12$}

Noting the expression for the mass ${\cal F}(Q,P)$ for $a=1,b^2\leq 1/2$ from \eqref{massfora=1} and the formula for the binding energy from \eqref{DelM}, it is very easy to see that
\bea
\text{when} && \quad {\cal F}(Q,P) = \sqrt{2}\,\Big( Q + \sqrt{1-2b^2}\,P \Big)\,, \nn\\
\Delta M &=& {\cal F}(Q_1+Q_2, P_1+P_2) - {\cal F}(Q_1,P_1) - {\cal F}(Q_2,P_2) = 0\,.
\eea
The vanishing of the binding energy correlates with the zero-force condition. 
Let us again note the similarity of this result with the vanishing of the binding energy for the BPS solution in the EMD theory for $a=1,b=0$. 
Finally, note that $\Delta M=0$ agrees with the vanishing of the curvature of the extremal curve, which in this case is just a straight line in the $u,v$ space.

\subsection{Binding energy when $a$ close to $1$, $b^2 = \ft12$}

In parallel to our discussion of the force, 
we wish to calculate the binding energy for a small perturbation of the case considered above.
Indeed, we may compute $\Delta M$
between two extremal black holes with charges $(Q_1,P_1)$ and $(Q_2,P_2)$ 
for the couplings $a=1+\epsilon,\, b^2=\ft12$, and investigate how it 
changes as we change $\epsilon$. Using the formula in \eqref{DelM}, 
the definition of $\cF(Q,P)$ in \eqref{defcF} and the relation between 
$Q$ and $P$ in \eqref{relQP}, we can write the expression for the binding
energy as
\bea
\fft{\Delta M}{\sqrt{8}} = (P_1+P_2)\,e^{a\,{\hat x}}\,f({\hat x}) - P_1\,e^{a\,x_1}\,f(x_1) - P_2\,e^{a\,x_2}\,f(x_2)\,,
\eea
where 
\bea
Q_1 = e^{2a\,x_1}\,P_1\,, \quad Q_2 = e^{2a\,x_2}\,P_1\,, 
\quad (Q_1 + Q_2) = e^{2a\,{\hat x}}\,(P_1 + P_2)\,.
\eea
The parameter ${\hat x}$ for the composite black hole is determined in terms of $x_1,x_2,P_1$ and $P_2$ by the equation,
\bea
(e^{2a{\hat x}} - e^{2ax_1})\,P_1 + (e^{2a{\hat x}} - e^{2ax_2})\,P_2 = 0\,.
\label{Pxeqn}
\eea
Without loss of generality, we shall assume $x_1>x_2$. From the equation above, we can then infer
\bea
x_1 > {\hat x} > x_2\,.
\eea
Instead of viewing $P_2$ as an independent parameter in the expression for
the binding energy, we can view it as being determined by eqn (\ref{Pxeqn}) 
in 
terms of $P_1,x_1,x_2$ and ${\hat x}$; that is, $\hat x$, instead of $P_2$, 
is now 
treated as an independent parameter. Finally, using the expression for 
$f(x)$ from \eqref{farounda=1} and the perturbation $w(x)$ from 
\eqref{wpert}, and expanding up to order $\epsilon$, the expression 
for $\Delta M$ becomes
\bea
\Delta M &=& \fft{P_1}{\sqrt{2}\,(e^{2\hat x} - e^{2x_2})}\,W(x_1, {\hat x}, x_2)\,\epsilon + {\cal O}(\epsilon^2)\,, \\
W(x_1,\hat{x},x_2) &=& (e^{2x_1} - e^{2x_2})\,\arctan e^{2 \hat{x}} + (e^{2\hat{x}} - e^{2x_1})\,\arctan e^{2x_2} + (e^{2x_2} - e^{2\hat{x}})\,\arctan e^{2x_1}\,. \nn
\eea
Note that the denominator above is positive since we assumed ${\hat x}>x_2$. We shall now prove that $W(x_1,\hat{x},x_2)$ is positive as well. To prove it, first let us note that, when we view $W(x_1,\hat{x},x_2)$ as a function of $\hat{x}$ with $x_1$ and $x_2$ fixed, then $W=0$ when $\hat{x} = x_1 = x_2$. Taking the derivative of $W$ with respect to $\hat{x}$ we find that it vanishes only at a single point in the range $x_1>\hat{x}>x_2$, and it corresponds to
\bea
\fft{dW(x_1,\hat{x},x_2)}{d\hat{x}} = 0 \quad \implies \quad e^{4\hat{x}} = \fft{(e^{2x_1}-\arctan e^{2x_1}) - (e^{2x_2}-\arctan e^{2x_2})}{\arctan e^{2x_1} - \arctan e^{2x_2}}\,.
\eea
In order to prove that $W$ is positive in the interval $x_1>\hat{x}>x_2$, 
all we need to show now is that the derivative 
$\dfft{dW(x_1,\hat{x},x_2)}{d\hat{x}}\Big\vert_{\hat{x}=x_2}$ 
is positive, i.e. the function $W$ rises from $\hat{x}=x_2$, reaches a positive maximum, decreases and becomes zero again at $\hat{x}=x_1$.

Defining 
$H(x_1,x_2) =\dfft{dW(x_1,\hat{x},x_2)}{d\hat{x}}\Big\vert_{\hat{x}=x_2}$,
one can view $H(x_1,x_2)$ as a function of $x_1$, keeping $x_2$ fixed. $H(x_1,x_2)$ has the property that $H(x_1,x_2)\vert_{x_1=x_2}=0$. Next, one can show that $H(x_1,x_2)$ has no turning point when $x_1>x_2$, i.e.
\bea
\fft{dH(x_1,x_2)}{dx_1} = 4\,e^{2(x_1+x_2)}\,\left(\fft{1}{1+e^{4x_2}} - \fft{1}{1+e^{4x_1}}\right) = 0 \quad \text{iff} \quad x_1 = x_2\,.
\eea
Finally it can be checked that
\bea
\fft{d^2H(x_1,x_2)}{dx_1^2}\Big\vert_{x_1=x_2}  = \fft{16\,e^{8x_2}}{(1+e^{4x_2})^2} > 0\,.
\eea
Hence we conclude that $H(x_1,x_2)>0$ when $x_1>x_2$. Therefore the sign of the derivative of $W$ at $\hat{x}=x_2$ is positive and as a consequence 
\bea
W(x_1,\hat{x},x_2) > 0\,, \quad x_1>\hat{x}>x_2\,.
\eea
Therefore when $a>1$, the binding energy $\Delta M$ is greater than zero.
Similarly, when $a<1$, $\Delta M$ is less than zero. 
Again we see that the sign of the force does not always correlate with the 
sign of the binding energy. In this case, the force correlates with 
the binding energy only when the initial black holes with charges 
$(Q_1,P_1)$ and $(Q_2,P_2)$ satisfy $Q_1Q_2<P_1P_2$.

\section{Conclusions}

In this paper we have extended our analysis of black holes and 
mass/force relations in EMD theories in \cite{crcvposa} to EMDA theories 
that involve an axion, described by the Lagrangian (\ref{EMDALag}).
In these theories, exact extremal, dyonic black hole solutions are not 
known for arbitrary values of the scalar couplings $a$ and $b$, but 
rather only for special cases.
Nonetheless, we succeeded in constructing a first order differential equation governing the behavior 
of the black hole mass, valid for arbitrary couplings $a$ and $b$.  
The benefit of this differential equation is that it enables the 
calculation of all the macroscopic properties of an extremal black hole, 
given any specific choice for $a$ and $b$. 
In the two special cases $a=b$ (with $a$ arbitrary) 
and $a=1,\,  b^2= \ft12$, we were 
able to solve the differential equation analytically, but not otherwise (although it is straightforward to solve numerically, doing so is beyond the scope of this paper).
For $a=b$ we also obtained an explicit solution for the black hole metric
itself, thanks to an enhanced $SL(2,\mathbb{R})$ symmetry of the 
equations of motion.

Beyond searching for new solutions in this theory, the primary motivation 
of this paper was to explore the extent to which one could identify any 
generic features in the behavior of the black hole mass or in the long-range 
forces between non-identical black holes (as well as their binding energy). 
Indeed, the main advantage of developing a general formalism for  
$f(x)$ and the extremal black hole mass is to give us an efficient way to 
calculate such forces. 
To date, there has been very little work on the structure of 
interactions between 
inequivalent black holes.  
In our previous study \cite{crcvposa} of the simpler EMD theory, we found an interesting correlation between the range of the dilatonic coupling $a$, the convexity/concavity of the extremal black hole mass (as a function of the charges of the black hole) and the sign of the long range forces and binding energy $\Delta M$.

In the setup of this paper, the presence of the axion significantly 
complicates the analysis, and the picture is not as clear.
Nonetheless, we were still able to identify certain regions of parameter space where the long range force $F_{12}$ between different black holes has a definite sign. 
In particular, we saw that when the couplings are equal to each other, $a=b$, the force is repulsive for $a^2>1$ and 
attractive for $0<a^2<\ft13$. In the remaining range $ \ft13\leq a^2<1$ 
it can have either sign depending on the black hole charges. 
We also found that when $a=1$ and $b^2 \leq \ft12$, the extremal black hole mass is linear in the electric and magnetic charges, much like for a BPS black hole,  and distinct solutions obey a no-force condition. 
This is peculiar, since we don't expect these solutions to be BPS. 
Thus, it would be interesting to understand the origin of the no-force condition, and why it is associated with this particular range of couplings. 
Moreover, a slight deviation from $a=1$ (keeping the axionic coupling in the 
range $b^2 \leq \ft12$) leads 
to either repulsive or attractive forces depending on whether $a$ is above 
or below 1, and on whether the electric charges of the solutions dominate over the magnetic ones.
Thus, even in the presence of the axion,  $a=1$ plays the role of a dividing line between 
attractive and repulsive interactions, just like in the EMD models we studied, with the main difference being that the size of the charges plays a role as well. 

Another feature we observed is that, unlike in the EMD theory, in certain regions of parameter space the sign of the long-range force does not agree with the sign of the binding energy $\Delta M$, which we recall is defined as in eq. (\ref{DeltaMdef}). For instance, for $a=b$ the binding energy is always negative 
(in agreement with the fact that the extremal mass curve is convex, as we 
explained in \cite{crcvposa}), 
while the force changes sign depending on the range of $a$, as we described 
above. We find a similar mismatch for $a$ close to 1 and 
$b^2=\ft12$, as a concrete example.  
This suggests that to determine whether the two black holes are likely to 
bind or not, one needs to consider
the interaction at short distances as well, and not rely exclusively on 
the asymptotic behavior.

Nonetheless, it is still interesting to ask why in certain ranges of parameter space $\Delta M$ and $F_{12}$ don't agree, while they do if we turn off the axion and restrict the theory to EMD. 
We wonder whether there may be additional constraints on the regions in parameter space associated with a mismatch between long-range forces and binding energies, which may rule them out altogether. 
Even if this were not the case, it is still valuable to understand the origin of this mismatch. 
One potential route to doing so is to generalize  \cite{Vinckers:2024tsa}, which used 
used the initial data analysis of \cite{Cvetic:2014vsa} to express the binding energy between distinct black holes in EMD theories in a way that can be directly related to the force. 
Repeating this analysis in the presence of an axion would let us identify the conditions that lead to 
the difference between asymptotic forces and binding energy,  
and might shed light on the structure of the black hole solutions themselves.
We wonder whether this may also shed light on constraints on low-energy EFTs and on the structure of axionic couplings.
We leave these questions to future work.

\section{Acknowledgments}

The work of C.N.P. is supported in part by the DOE Grant No. DE-SC0010813.
The work of M.C. is supported by the DOE (HEP) Award DE-SC0013528, the Simons Foundation Collaboration grant 724069 on ``Special Holonomy in Geometry, Analysis and Physics", the Slovenian Research Agency (ARRS No. P1-0306) and the Fay R. and Eugene L. Langberg Endowed Chair funds.
The work of S.C. was supported in part by the National Science Foundation under Grant No. PHY-2210271. 
This research was supported in part by the National Science Foundation under Grant No. NSF PHY-1748958. A.S. would like to thank ICTP, Italy for supporting the Feza Gursey Center via the Affiliated Centers Programme. The authors would like to thank Cook's Branch 2024, where part of this work was carried on, for the hospitality and support.

\section*{Appendix: EMDA Generalisation of Non-Extremal Gibbons-Maeda Black Holes \label{apdx A}}

  For completeness, we record here the generalisations of the non-extremal
Gibbons-Maeda black holes to the case of the EMDA theory with $b=a$, again
obtained by making use of the $U(1)\subset SL(2,\R)$ global symmetry
of the $b=a$ theories.  The non-extremal Gibbons-Maeda black holes 
in the EMD theory are given by \cite{gibmae,cvegibpop}
\bea
ds^2 &=& -\Delta\, dt^2 + \fft{dr^2}{\Delta} + R^2\, d\Omega^2\,,\nn\\
e^{a\phi} &=& f_-^{\fft{2a^2}{1+a^2}}\,,\qquad A=-P\, \cos\theta\, d\varphi\,,
\nn\\
\Delta&=& f_+\, f_-^{\fft{1-a^2}{1+a^2}}\,,\qquad R^2 = r^2\, 
   f_-^{\fft{2a^2}{1+a^2}}\,,\qquad f_\pm = 1 -\fft{r_\pm}{r}\,,
\eea
with $P^2= \dfft{4 r_+\, r_-}{1+a^2}$.  

The parameters $r_\pm$ are related to $\beta,\gamma$ that appear in \eqref{GibMae1} as $r_\pm = \ft12(\beta \pm \gamma)$. Applying the $U(1)$ transformation, we
obtain non-extremal static dyonic solutions in the $b=a$ EMDA theory, with
\bea
ds^2 &=& -\Delta\, dt^2 + \fft{dr^2}{\Delta} + R^2\, d\Omega^2\,,\nn\\
e^{a\phi} &=& \fft{(P^2+Q^2)\, h}{P^2+Q^2\, h^2}\,,
\qquad  \chi= -\fft{P Q (1-h^2)}{a\, (P^2+Q^2\, h^2)}\,,
  \qquad A= \fft{Q}{r}\, dt -P\, \cos\theta\, d\varphi\,,
\nn\\
\Delta&=& f_+\, f_-^{\fft{1-a^2}{1+a^2}}\,,\qquad R^2 = r^2\,
   f_-^{\fft{2a^2}{1+a^2}}\,,\qquad h= f_-^{\fft{2a^2}{1+a^2}}
   \,,\qquad f_\pm = 1 -\fft{r_\pm}{r}\,,
\eea
and with
\bea
P^2+Q^2= \fft{4 r_+\, r_-}{1+a^2}\,.
\eea

\end{document}